\documentstyle[aps,prl,psfig]{revtex}
\begin{document}
\title{Robust test for detecting non-stationarity in data 
from gravitational wave detectors}
\author{Soumya D. Mohanty\cite{email}}
\address{Center for Gravitational Physics and Geometry, \\ Pennsylvania State 
University, University Park, Pa 16801, U.S.A.}
\date{Received 24 January 2000; published 24 May 2000}
\maketitle
\begin{abstract}
It is difficult to choose detection thresholds for tests of non-stationarity
that assume {\em a priori} a noise model  if the data is statistically
uncharacterized  to begin with. This is a potentially serious problem
when an automated analysis is required, as would
be the case for the huge data sets that
large interferometric gravitational wave detectors will produce.
 A solution is proposed in the 
form of a {\em robust} time-frequency test for detecting non-stationarity 
whose threshold for a specified false alarm rate is almost independent
of the statistical nature of the ambient stationary noise. 
The efficiency of this test in detecting bursts is compared with that of an 
ideal test that requires prior information about both the statistical 
distribution of the noise and also the frequency band of the
burst. When supplemented with an approximate knowledge of the
burst duration, this test can detect, at the same false alarm
rate and detection probability, bursts that are about 3 times larger
in amplitude than those that the ideal test can detect.
Apart from being robust, this test has properties which make 
it suitable as an online monitor of stationarity.

\end{abstract}
\section{Introduction}

Each of the large interferometric gravitational Wave detectors
that are now under construction~(LIGO~\cite{LIGO},
VIRGO~\cite{VIRGO}, GEO~\cite{GEO}, TAMA~\cite{TAMA}) will produce
a flood of data when they come online in a few
years. Apart from the ``main'' data channel carrying
 measurement of strain in the arm lengths, there will
be a few hundred auxiliary channels~\cite{channel_list} at each site
associated with system and environmental monitors,
such as seismometers and magnetometers. Their role
would be to monitor the state of the detector and its
environment so that any unusual event in the main
channel or an unexpected behavior of the detector
can be diagnosed properly. (The sum total of  raw
data from the LIGO detectors will be produced at the
rate of $\sim 10$ megabytes~\cite{ldas} every second.)

Under ideal conditions, each data channel
would carry  stationary noise. For the main
channel, this would reflect a steady state
of the interferometer and for the auxiliary
channels, a steady state of the environment.
However, experience with prototypes as well
as with the several resonant mass detectors that have
been operating for quite some time shows that
this situation does not hold in reality. 
There will always be episodes of non-stationarity
though their rates and durations will depend 
on the choice of the detector site and other factors.

Detecting non-stationarity is important both in the 
main channel, because some non-stationarity could be
of astrophysical origin, and also in the auxiliary
channels where it can be an important diagnostic of the instrument
or its environment. It is also important when estimating a statistical model
of the detector noise where it is essential that the data segment used
be stationary. [The deleterious effects of non-stationarity
on power spectral density~(PSD) 
estimation were noted in~\cite{allen+etal:40m}.]

Several  methods for detecting non-stationarity 
that are relevant in this context
have already been considered in the gravitational wave data 
analysis literature~\cite{arnaud+etal:99,flannagan+hughes}.
 However,  these methods share an unsatisfactory feature which is that
 the computation of the detection threshold corresponding to
a specified {\em false alarm rate} requires an {\em a priori}
knowledge of a statistical model of the stationary ambient noise.
An error in the model leads to an error in
our knowledge of the false alarm rate. 
In the real world such prior models are usually not
available and it is necessary to estimate noise models from the data 
itself. Even if a model exists, it will almost always 
have some free parameters (the  variance being a trivial
example) whose values would have to be estimated from the data
fairly regularly, especially
 in the case of a complicated instrument such as a laser interferometer
or its environment monitors. 

Thus, when confronted with an uncharacterized dataset,
 an experimenter who is only limited to methods such as the
above can face considerable uncertainty in 
fixing a threshold for the test before analyzing the data.
 For a sufficiently
small dataset, the analyst can start with {\em ad hoc}
thresholds and work in some iterative sense towards
 a statistically satisfactory 
conclusion.  The
problem becomes more serious when the data set to be
analyzed is so large that it becomes necessary to substantially automate
the analysis, as would most certainly be needed in the case of
the large interferometers.  An additional set of problems will
arise when analyzing auxiliary channels since ambient terrestrial noise may
be intrinsically more difficult to characterize and  
 have a variable nature.

We introduce here, in the context of gravitational wave data analysis,
 a test for detecting non-stationarity
 for which the issue of fixing the correct 
threshold is trivial by design.
The false
alarm rate for such a {\em robust} test depends weakly
 on the statistics of the ambient noise and is
 specified almost completely by the detection
threshold alone. 
In the present paper we concentrate on 
short duration non-stationarity or {\em bursts} since they
are likely to be the most common types of non-stationarity
in gravitational wave detectors. We find that the robustness 
of the test improves
for smaller false alarm rates, which is precisely
the regime of interest.  If required,
the test can be optimized in terms of the 
duration of the bursts that need to be detected. 

We compare the efficiency of
this test in detecting narrowband bursts with  that of an ideal test which
requires  both a noise model and prior knowledge of
the frequency band~(center frequency and bandwidth)
 in which the bursts occur.
We find that supplementing our test with an 
approximate prior knowledge of the burst duration allows
it to detect, at the same false alarm rate and detection
probability, bursts with a peak amplitude 
that is a factor of $\sim 3$ larger than that of the bursts which
the ideal test can detect.

Apart from being robust, it also has the following properties
that make it useful as an online monitor of stationarity.
 The computational cost associated with this test is quite
small.
Areas of non-stationarity are clearly distinguished, 
in the time-frequency
plane, from areas of stationarity. Apart from making the 
output simple
to understand visually, this will allow an automated
routine to catalogue burst information such as the time of 
occurrence and frequency band. 
 
The detection of non-stationarity has been actively studied in Statistics
for quite some time~\cite{nonstat} and numerous tests suitable for
a wide variety of non-stationary effects exist in the literature.
The central idea behind our test is the detection of statistically
significant {\em changes} in the PSD.
As a means of detecting
non-stationarity, this idea is quite natural and has been proposed 
in several earlier works. (See, for instance,\cite{priestley,grasp}.)
though what constitutes a change and how it is measured can
be defined  in many different ways leading
to tests that differ  statistically as well as computationally.
  The specific implementation 
presented in this paper leads to a statistically robust test.
 The issue of  robust
tests for non-stationarity, though important as we have argued,
 has not been considered in  gravitational wave detection so far.
The same concerns as well as a more rigorous treatment exist in the 
Statistical literature~\cite{brodsky+darkhovsky}. Our present
 work was, however, done independently and this test is a new
contribution.

The paper is organized as follows. In Section~\ref{sec1} we formally state the
problem addressed in this paper. 
Section~\ref{sec2} describes the Student {\em t}-test which lies
at the core of our test. This is followed by a 
discussion of the basic ideas that lead to the test
and why the test can be expected to be robust. 
In Section~\ref{sec3}, the test is characterized statistically in term of
its false alarm rate and detection power. The main results of this 
paper are also presented in this section. 
The computational cost associated with this test is
discussed in Section~\ref{sec3_4}.
This is followed by our conclusions and pointers to future work in 
Section~\ref{sec5}.
\section{Formal statement of the problem}
\label{sec1}

A random process $x(t)$ is said to be {\em 
strictly stationary}~\cite{Goodman} if
the joint probability density $P(x(t_i),x(t_i+\delta_1),x(t_i+\delta_2),
\ldots,x(t_i+\delta_n))$
of any finite number, $n$, of samples is independent of $t_i$.
Often, one uses a less restrictive definition called {\em wide sense 
stationarity} which demands only that the mean ${\rm E}\left[x(t_i)\right]$ and
the autocovariance ${\rm E}[(x(t_i)-{\rm E}[x(t_i)])( x(t_i+\tau)
-{\rm E}[x(t_i+\tau)])]$ be independent of $t_i$.
A random process not satisfying any of the above definitions is 
called {\em non-stationary}.

We assume that the ambient noise 
in the data channel of interest is wide sense stationary
over sufficiently long time scales and a burst  is an
episode of non-stationarity with a much smaller duration.
 That is,
the occurrence of a burst lasting from $t=t_0$ to $t=t_1$ in a segment
$x(t)$ of data ($0\leq t\leq T$) means that
\begin{equation}
x(t) = \left\{ \begin{array}{lr}
\mbox{wide sense stationary} & 0\leq t\leq t_0\\
\mbox{non-stationary} & t_0 \leq t\leq t_1\\
\mbox{wide sense stationary} &  t_1 \leq t\leq T
\end{array}
\right. \;,
\end{equation}
where $t_1-t_0 \ll T$.
In practice, only a {\em time series} ${\bf x}$
consisting of regularly spaced samples of $x(t)$ is available instead of
$x(t)$ itself.
Thus, given the time series ${\bf x}$, we want to decide between 
the following two hypotheses about ${\bf x}$ :
\begin{enumerate}
\item {\em Null Hypothesis} $H_0$ : ${\bf x}$ 
is obtained from a wide sense stationary random process.
\item {\em Alternative Hypothesis} $H_1$ : ${\bf x}$
is obtained from a non-stationary random process.
\end{enumerate}
 The 
{\em frequentist approach}~\cite{Kendall:v2} to this 
decision problem, which is 
followed here,  begins by
 constructing  a function
${\cal T}({\bf x})$, called a {\em test statistic},
 of the data ${\bf x}$. If the data ${\bf x}$ is such that ${\cal T}({\bf x}) \geq \eta$, for 
some threshold $\eta$, the null hypothesis is rejected in favor of the 
alternative hypothesis for that ${\bf x}$.

Since ${\bf x}$ is obtained from a random process, there exists
a finite probability,  that ${\cal T}({\bf x})$ crosses the 
threshold even when the data is stationary. Such an event is 
called a {\em false alarm} and the rate of such events 
over a sequence of data ${\bf x}$ is called the {\em false alarm rate}.
The threshold $\eta$ is determined by specifying the
 false alarm rate that the analyst is willing to tolerate.

To compute the threshold, we need to know the distribution
function of ${\cal T}({\bf x})$ when $H_0$ is true. 
This distribution can, in principle,
 be obtained if  the joint distribution of ${\bf x}$~(i.e., a noise model)
 is known.
However, as mentioned in the introduction, such prior
 knowledge is usually incomplete, if it exists at all,
in the real world.
The only solution then is to estimate the joint distribution
 from the data itself. Therefore, one must first find a 
stationary segment of the data, by detecting and then rejecting
non-stationary parts,  but that brings us back to our primary 
objective itself!

To get around this paradox, we must {\em construct}
 ${\cal T}({\bf x})$ such that
its distribution
 is as independent as possible of the distribution of the
 data under the null hypothesis.
If the distribution of the test statistic 
is strictly independent of the distribution of ${\bf x}$, the test is called~\cite{nonpar}
{\em non-parametric}. If the test statistic distribution depends on
the distribution of ${\bf x}$ but only {\em weakly}, the test is 
said to be a {\em robust } test. Tests which do not have either of
these properties are called {\em parametric}.
Formally, therefore, the aim of this work is to find a non-parametric,
 or at least  a robust test, for non-stationarity.

\section{Description of the test}
\label{sec2}
\subsection{Student's $t$-test}
\label{sec2_1}
Before we describe our test for non-stationarity, it is best to
discuss Student's {\em t}-test~\cite{nonpar} in some detail since this standard statistical
test plays an important role in what follows. 

Student's {\em t}-test is designed to address the following problem.
Given a set of $N$ samples, $\left\{ x_1,\ldots,x_N\right\}$,
 drawn from a Gaussian distribution of unknown mean  and variance, how do we 
check that the mean $\mu$ of the distribution is non-zero? 
In Student's {\em t}-test, a test statistic $t$ is constructed,
\begin{equation}
t  =  \frac{\widehat{\mu}\sqrt{N}}{\sqrt{\widehat{s}^2}}\;,
\end{equation}
where
\begin{eqnarray}
\widehat{\mu} &:=& \frac{1}{N} \sum_{j=1}^N x_j\;,\\
\widehat{s}^2 & := & \frac{1}{N-1} \sum_{j=1}^N \left(x_j-\widehat{\mu}\right)^2
\;.
\end{eqnarray}
The distribution of $t$ is known~\cite{univariate}, both when $\mu=0$
and $\mu\neq 0$.
To check whether $\mu \neq 0$, a two-sided threshold is set on
{\em t} corresponding to a specified false alarm 
probability. If {\em t} crosses the threshold on either side, the null
hypothesis $\mu=0$ is rejected in favour of the 
alternative hypothesis $\mu\neq 0$.

Of interest to us here are two main properties of the {\em t}-test. 
First, if  two sets of independent samples $X=\left\{x_1,\ldots,
x_N\right\}$ and $Y = \left\{y_1,\ldots,y_N\right\}$  are
drawn from Gaussian distributions with the same but unknown variances, the 
{\em t}-test can be employed to check whether the means of the two 
distributions are equal or not. This can be done simply by constructing 
a third set of samples $Z=\left\{y_1-x_1,\ldots,y_N-x_N\right\}$, which would again be 
Gaussian distributed, and then testing, as shown above, whether the mean of the distribution
from which $Z$ is drawn is non-zero or not. 

The second important property~\cite{geary:36} of the {\em t}-test is its {\em robustness} :
As long as  the underlying distributions from which the two samples are 
drawn are identical, but
not necessarily Gaussian, the distribution of the {\em t} statistic does
not deviate much from the Gaussian case. The lowest order corrections to the 
mean and variance of the distribution being ${\cal O}(N^{-5/2})$ and 
${\cal O}(-N^2)$ respectively.

\subsection{An outline of the test}
\label{sec2_2}

We present here an outline of our test. 
The details of the actual algorithm are presented 
in the appendix.

From Sec.~\ref{sec1}, it is clear that a 
direct signature of non-stationarity is
 a change in the autocovariance function. 
This implies that the PSD
of the random process should also change since the 
it is the Fourier transform 
of the autocovariance function~\cite{Goodman}.
Therefore, the basic idea behind our test
is the detection of a change in the PSD of a time series. 

\begin{figure}
\centerline{\psfig{file=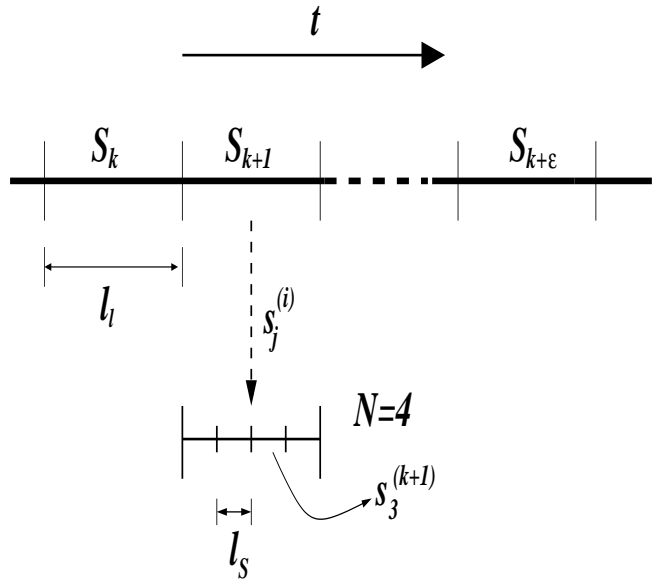,height=3in,width=3.4in}}
\caption{\label{figx1} Schematic
of the data stream subdivisions.
In this example, we have chosen $N=4$. (Thus, $j=1,2,3,4$.)}
\end{figure}

The test 
involves the following steps (see Fig.~\ref{figx1} also).
\begin{enumerate}
\item The time series to be analyzed is divided into adjacent but 
disjoint segments
of equal duration $l_l$. \label{outline_step1}
\item Take two such disjoint data segments $S_k$ and $S_{k+\epsilon}$ 
separated by a time interval $(\epsilon-1) \, l_l$, $\epsilon =1, \, 2,\,\ldots$.
We would like to compare the PSDs of these two segments and test if there
is a significant difference.\label{outline_step2}

\item Subdivide each of the two segments into $N$ 
subsegments of equal duration. Thus, segment
$S_i$, $i\in\left\{ k,\, k+\epsilon\right\}$, gives 
us $N$ subsegments, each of duration 
$l_s=l_l/N$, which we denote by $ s^{(i)}_j$, $j = 1,\, 2, \ldots, N$. 
This is an intermediate
step in the estimation of the PSD of each segment $S_i$.\label{outline_step3}

\item Compute the {\em periodogram} of each $s^{(i)}_j$. 
A periodogram is the squared modulus of the Discrete
Fourier Transform~(DFT) of a time series~[Eq.~(\ref{pdg}]. 
\label{outline_step4}

\item For every frequency bin, therefore, we obtain a set  $X$ of
$N$ numbers
from $S_k$ and similarly another set $Y$ from $S_{k+\epsilon}$. 
In a conventional 
estimation of the PSD of a segment, say $S_k$, we would simply average the
corresponding set $X$. However, since we want to {\em compare} 
two PSDs, we do the
following instead.\label{outline_step5}

\item Perform Student's {\em t}-test for equality of mean on these 
two independent sets 
 $X$ and $Y$. If the {\em t} statistic crosses a preset threshold 
$\eta$, then
a significant change in the mean is indicated, otherwise not.
\label{outline_step6}

\item Repeat step~\ref{outline_step6} for all frequency bins 
in exactly the same 
manner. \label{outline_step7}
\end{enumerate}
Steps~\ref{outline_step2} to~\ref{outline_step7}
should then be repeated with another pair of disjoint segments
 $S_{k+1}$ and $S_{k+\epsilon+1}$ and so on.

\begin{figure}
\centerline{\psfig{file=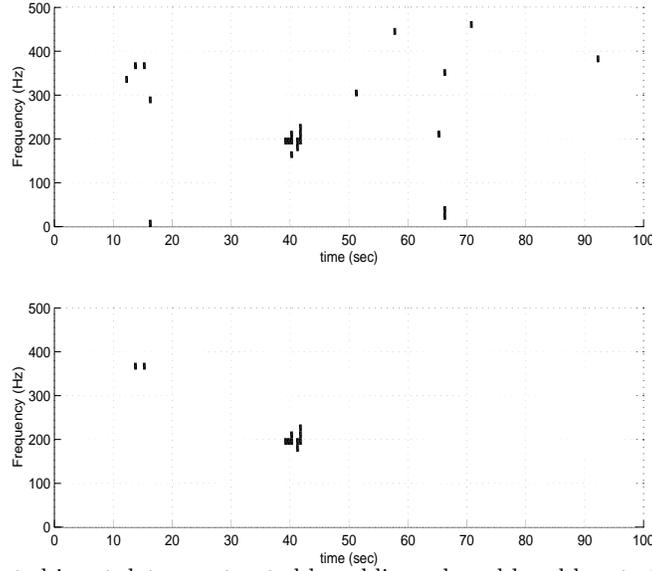,height=3in,width=3.4in}}
\caption{\label{fig1} Test output for simulated  input data constructed by
 adding a broad band burst of effective duration 1.0~sec, center frequency
200~Hz and bandwidth 200~Hz to stationary white Gaussian noise.
The sampling frequency was 
chosen to be 1000~Hz. (a) Top:  raw image obtained
after applying the {\em t}-test threshold. (b) Bottom: 
result of applying the veto to the
image in (a). The cluster that occurs earlier in (b) is a false event
while the next cluster corresponds to the burst.}
\end{figure}

Thus, the output of the test at this stage  is a two dimensional image with time along one axis 
and frequency along the other. In this image, every frequency 
bin for which the threshold $\eta$ is crossed can be thought of 
as being colored black while the remaining are colored white.
Hence, white areas in this image would indicate stationarity while 
the contrary would be indicated by the black areas. A sample 
image is shown in Fig.~\ref{fig1}(a). It is the result of applying the
test to a simulated time series constructed by adding a 
broad band burst to stationary white Gaussian noise 
(see Sec.~\ref{sec3_1} for 
definitions).

Not all black areas would, however, correspond to non-stationarity. Most
of them would be random threshold crossings caused
by the stationary noise itself. We search, therefore, 
for {\em clusters} of black pixels in the image 
which pass a veto that can be motivated as follows. Suppose the burst
 is fully contained in one segment, say, $S_k$. Then one would 
expect the {\em t}-test threshold to be crossed once when  comparing $S_k$ with
$S_{k-\epsilon}$ and again when $S_k$ is compared with $S_{k +\epsilon}$.
This leads to a characteristic ``double bang''
structure for the cluster of black pixels. We throw away all other groups
of black pixels that do not show such a feature.
(This scheme is defined rigorously in the appendix.)
Fig~\ref{fig1}(b) shows the result obtained by applying this
veto to the image in Fig~\ref{fig1}(a). One of the 
cluster is at the location of the added burst while the other
is a false event.

\subsection{Why is this test robust?}
\label{sec2_3}

This test can be expected to be robust for two reasons.
First, the periodogram at any frequency is asymptotically 
exponentially~\cite{exp_asymp,exponential}
distributed. This can be 
heuristically explained as follows. The DFT of a time series
is a linear transform. If the number of time samples
 in a random time series is sufficiently large, it then
follows from the {\em central limit theorem} that the DFT of that time series
 will have, at each frequency, imaginary and real parts which are
distributed as Gaussians. Since the basis functions used
in a DFT are orthogonal, the real and imaginary parts also
tend towards being statistically independent. This implies
 that, for a sufficiently large number of time samples,
 the periodogram, which is 
simply the squared modulus of the DFT,  is exponentially
distributed at each frequency.

The second reason which should make the test 
robust is the fact, mentioned earlier, that the {\em t}-test is 
robust against non-Gaussianity when the two samples being
compared have identical distributions. Under the 
null hypothesis of stationarity, we do indeed
 have identically distributed sets in our case.

Since the asymptotic distribution of a periodogram is 
independent of the statistical distribution of the time samples, much of the
  information about  the time domain statistical distribution is lost in
the frequency domain. Thus, the {\em t}-test ``sees''  nearly
exponentially distributed samples whereas the time domain samples
may have a Gaussian or Non-Gaussian distribution. 
Added to this, the robustness of the {\em t}-test also removes
information about the time domain statistical distribution.
Further,  the {\em t}-test checks for a {\em change} in the
mean value and is insensitive to the absolute value of the mean.
This is strictly true in the Gaussian case but, because of the 
robustness of the {\em t}-test, it should also hold to a large
extent for the exponential case.

These basic considerations suggest strongly that the
test as a whole should be robust. However, the test
also involves some other steps beyond just a simple {\em t}-test.
First, the same segment is involved twice in a 
{\em t}-test~(c.f.~Sec.~\ref{sec2_2}).
 Thus, for any $k$, samples $k$ and $k+\epsilon$ in
 the sequence of {\em t} values at a given frequency will be correlated to 
a large extent. Second, we impose a non-trivial veto.

The above features of the test, though well motivated and conceptually
simple, make a straightforward
analytical study of the test difficult. Therefore, to establish the robust
nature of the test and quantify its performance,  we must  
follow a more empirical approach based on Monte Carlo simulations. This is 
the subject of the next section. An analytical treatment of the test is currently 
under development.

\section{Statistical characterization of the test}
\label{sec3}

Our main aim in this section
is to demonstrate the robust nature of the test
and to study the efficacy of this test in detecting non-stationarity.
Since, we need to use Monte Carlo simulations for understanding
these statistical aspects of the test, we discuss only a few
selected cases in this paper.

For a test to qualify as robust the 
threshold should be almost completely specified by the false
alarm rate without requiring any assumptions about the statistics of the 
data.  The false alarm rate,
 in the context of this test,  is the rate at which 
clusters of black pixels occur when the input to the test 
is a {\em stationary} data stream. 
To obtain the false alarm rate, several realizations
of {\em stationary} noise are generated and the test is 
performed on each. For a given threshold, 
the number of clusters detected over all the realizations  
provides an estimate of the false alarm rate at that threshold. 

The efficacy of a test in detecting a deviation from the null
hypothesis is measured by the {\em detection probability} of 
the deviation. In this paper, we measure
the detection probability of different types of {\em bursts}
that appear additively in stationary ambient noise.
Realizations of signals from a fixed class~(such as
narrowband or broadband bursts of noise)  are generated, to each of which
we add a realization of stationary noise.  The test is
applied to the total data and we check whether a cluster of
black pixel appears in a specified area of the time-frequency plane.
This fixed area, which we call the {\em detection region}, is specified in advance of the simulation.
The ratio of the number of realizations having a cluster in the 
specified area to the total number of realizations gives
an estimate of the detection probability for bursts of that class.

The function that maps the test threshold into false alarm rate 
depends on the test parameters, $l_l$, $l_s$ and
 $\epsilon$~(c.f.~Sec.~\ref{sec2_2}).
 Therefore, for each choice of these test parameters,
the  test must be calibrated separately
using a Monte Carlo simulation. However, thanks to the robust 
nature of the test, the simulation needs to   be performed only once and
for a simple noise process such as Gaussian white noise
which need not have any relation to the actual random process
at hand. 
The role of the test parameters is discussed  in more detail
in Sec.~\ref{sec3_3}.

\subsection{False alarm probability} \label{fprob_sec}
\label{sec3_1}
We perform a Monte Carlo
simulation for each of the representative cases below
and show that the false alarm rate, as a function of threshold,
is the same for all of them. 

Each realization of the input 
data is a 10~sec long time series and each simulation
uses $5000$ such realizations.
We can look upon all
the separate realizations of the input as forming parts of a single
data stream~($5000\times 10$~sec long) and, if we assume that
false alarms occur as a Poisson process, the false alarm rate~(in number of
events per hour) is
given by the total number of false alarms over all realizations
divided by $5\times10^4/3600$.

The various cases considered here are as follows.

{\em (i) White Gaussian noise} ($\sigma=1$)---  The time series
 consists of independent and identically distributed
Gaussian random variables.
The standard deviation $\sigma$ of the Gaussian random variables is unity and their mean is zero.

{\em (ii) White Gaussian noise} ($\sigma=10$) --- Same as above but 
with $\sigma=10$.

{\em (iiI) White non-Gaussian noise ---} All details in this simulation
are the same as above except that the distribution of each sample is
now chosen to be an exponential with $\sigma=1$. 

{\em (iv) Colored noise ---} We generated Gaussian, zero mean noise  
with a PSD as shown in Fig.~\ref{fig2}.  
The overall normalization is arbitrary but the 
noise is scaled in the time domain to make its variance unity.
 This PSD was derived from the expected initial LIGO PSD, as provided 
in~{\protect \cite{ligo_psd}}, by truncating the latter below 5~Hz 
and above 800~Hz followed by the application of
 a band pass filter with unity gain between 50~Hz and 500~Hz.

The range covered by the above types of statistical models 
is much more extensive than would be required in practice.
By applying the test to such extreme situations,
we can {\em bound} the variations in the false alarm 
rate versus threshold curve that would occur
in a more realistic situation. 
In considering this range of models for the stationary
background noise, we have gone from a two-sided
distribution to a completely one side distribution. The output from
most channels would be two sided and, hence, closer to a 
Gaussian than the Exponential distribution considered here.

The results are shown in Fig.~\ref{fig3}, \ref{fig4} and~\ref{fig5}. 
For the small false alarm rates~($< 5$/hour) that will be required in practice,
the test is clearly shown to be 
very insensitive to the statistical nature of the data. The largest variation
is between the Gaussian and exponential case while there is hardly
any variation, even at large false alarm rates,   among the Gaussian
cases.  The variation between the Gaussian and exponential case
is less than $\sim 50\%$ in the worst case. As explained above,
this should be treated
as an upper bound on the error one might expect in practice.

Figs.~\ref{fig3}, \ref{fig4} and~\ref{fig5} correspond to  different sets
 of test parameter values. 
The  threshold for a given false alarm rate does depend, as one
may expect, on the parameters of the test $l_s$, $l_l$ and $\epsilon$.
 Because of the robust nature, however, given a particular set of
 parameter values only  a single Monte Carlo simulation has to be performed with, say,
white Gaussian
noise, in order to obtain the corresponding false alarm rate versus
 threshold curve. 

The parameter values for Fig.~\ref{fig3} were chosen
to be the same as those that will be used in the following
section. We also consider in that section the case of 
a band pass filtered and down sampled time series.
Fig.~\ref{fig5} uses parameter values appropriate to the latter
while the choice for Fig.~\ref{fig4}  is explained in more
detail in Sec.~\ref{sec3_3}.

\subsection{Detection probability}
\label{sec3_2}

A burst has 
an effectively finite duration and is itself an instance of 
a stochastic process.  
We consider the following combinations of background noise, bursts and test parameters
$l_l$, $l_s$ and $\epsilon$. 
The 
sampling frequency of the data is assumed to be $1000$~Hz.

\begin{figure}
\centerline{\psfig{file=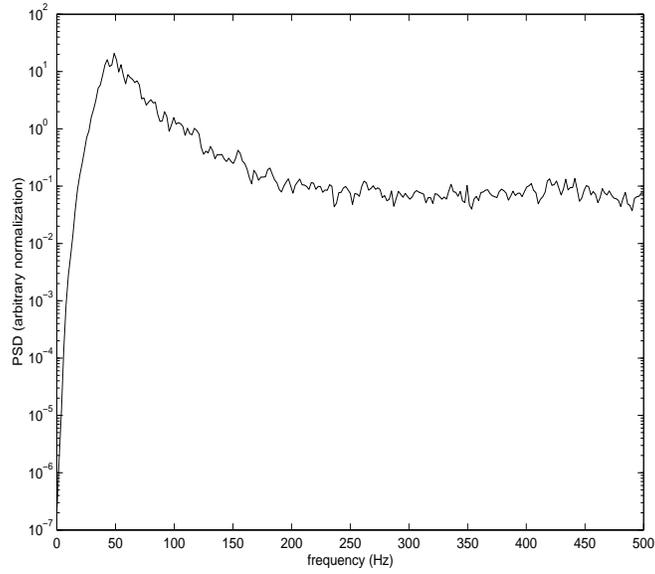,height=3in,width=3.4in}}
\caption{\label{fig2}PSD for the colored Gaussian noise
used in this paper.}
\end{figure}

\begin{figure}
\centerline{\psfig{file=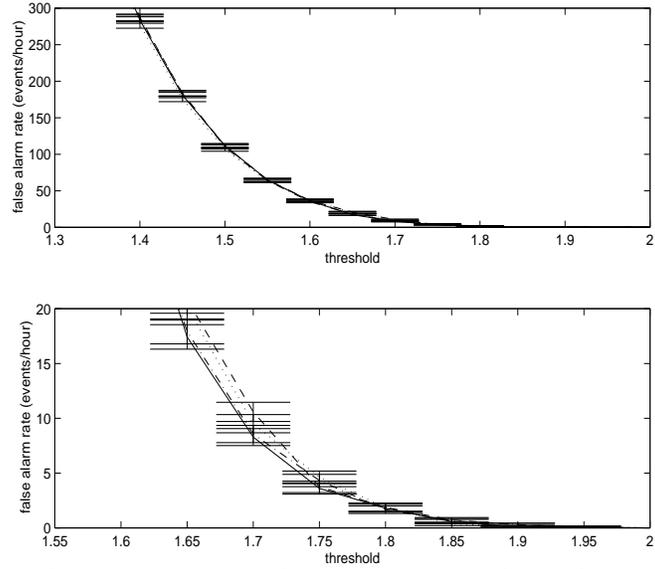,width=3.4in,height=3in}}
\caption{ \label{fig3} False alarm rate as function of threshold for 
different types of stationary input noise. The sampling frequency of the input is 
1000~Hz. Bottom panel: zoomed in view of the top panel. 
Solid line: white Gaussian
noise ($\sigma=1$). Dashed 
line: white Gaussian
noise ($\sigma=10$). Dotted line: white 
exponential noise ($\sigma=1$).
Dash-dotted line: colored Gaussian noise. 
The error bars correspond to $1\sigma$ deviations. 
The test parameters values are
$l_l=0.5$~sec, $l_s=0.064$~sec and $\epsilon=3$.}
\end{figure}

\begin{figure}
\centerline{\psfig{file=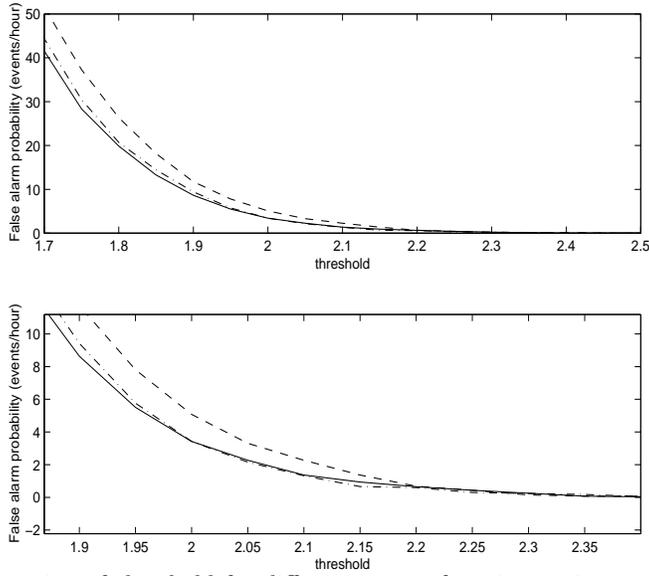,width=3.4in,height=3in}}
\caption{\label{fig4} False alarm rate as function of threshold for 
different types of stationary input noise. The sampling frequency of the input is
$1000$~Hz. Bottom panel: 
zoomed in view of the top panel.  Solid line: 
white Gaussian  noise ($\sigma=10$). Dashed line:
  white exponential noise ($\sigma=1.0$).
 Dash-dotted line: colored Gaussian noise.
The test parameters values are
$l_l=1.25$~sec, $l_s=0.064$~sec and $\epsilon=3$.}
\end{figure}

\begin{figure}
\centerline{\psfig{file=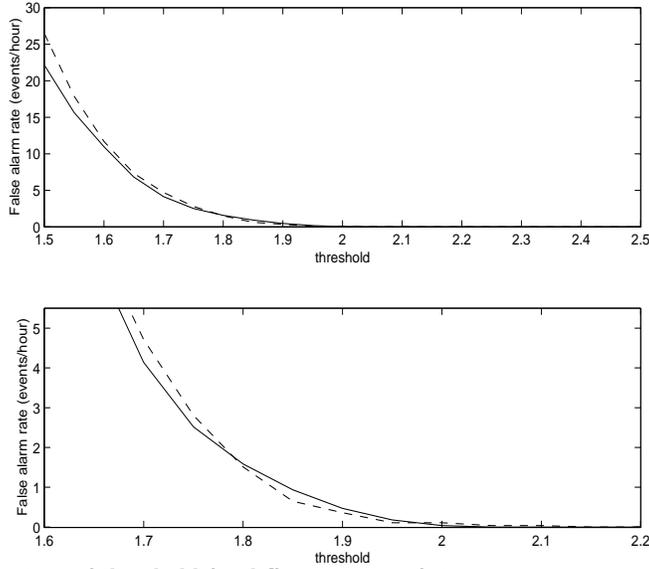,width=3.4in,height=3in}}
\caption{\label{fig5} False alarm rate as  function of threshold for 
different types of stationary input noise. The sampling frequency of the 
input is 40~Hz. Bottom panel: 
 zoomed in view of the top panel.  Solid line: white
 Gaussian white noise with ($\sigma=10$).
 Dashed line: white exponential noise ($\sigma=1.0$).
The test parameters values are
$l_l=1.0$~sec, $l_s=0.1$~sec and $\epsilon=3$.}
\end{figure}

The background noise is a zero mean stationary Gaussian
process with a PSD that matches the expected initial
 LIGO PSD (c.f.~Fig.~\ref{fig2}).
The burst is a {\em narrow band } burst constructed by band pass filtering a white
Gaussian noise sequence followed by multiplication of the filtered
output with a time domain window. Let the width of the pass band be $W$ and 
its central frequency be $f_c$. The time domain
window function is chosen to be a Gaussian 
in shape ($\exp\left(-t^2/2\Sigma^2\right)$) where
$\Sigma$ is chosen such that when $t=0.5$~sec, the 
window amplitude drops to 10\% of its maximum value (which is unity at $t=0$).
The burst has, therefore, an effective duration of $\sim 1$~sec.
After windowing, the peak amplitude of the
burst is normalized to a specified value.
The test parameters are $l_l=0.5$~sec, $l_s=0.064$~sec  and $\epsilon=3$.
(
$l_s=0.064$~sec corresponds to 64 points, a power of 2, in order to 
optimize the Fast Fourier Transforms needed for computing the
periodogram for each subsegment.) 

We consider two types of narrow band bursts. Type (1) has $f_c=200$~Hz, while
type (2) has $f_c=100$~Hz. $W=20$~Hz for both types of bursts.
The detection region, which is the area in the time frequency plane
that must contain a cluster of black pixel for a valid detection, is
chosen in both cases to be $1.0$~sec and $80$~Hz wide
in time and frequency respectively. It is centered at  the location of
the window maximum in time and at $f_c$ in frequency.

For each type of burst, we empirically determine
the peak amplitude required  in order for the burst to have 
 a detection probability of $\simeq 0.8$.  This is done 
at several different values of the detection threshold 
  corresponding to false alarm rates of 
1 false event in 1/2, 1, 2, or 3 hours.
The results are tabulated in Table~\ref{tableI}.

As shown later in Sec.~\ref{sec3_3},
the above choice for the test parameters, especially the value of $l_l$,
optimizes the test for detecting bursts which effectively last for 
$\sim 1$~sec. 
We have, therefore, presented the best performance
the test can deliver for detecting bursts with this duration. 
 Note that the same set of parameters optimize the test for
detecting bursts that occur in very different frequency bands.
Thus, the duration of a burst is effectively the only characteristic
that needs to be considered when optimizing the test.
This point is discussed further in Sec.~\ref{sec3_3}.

\begin{figure}
\centerline{\psfig{file=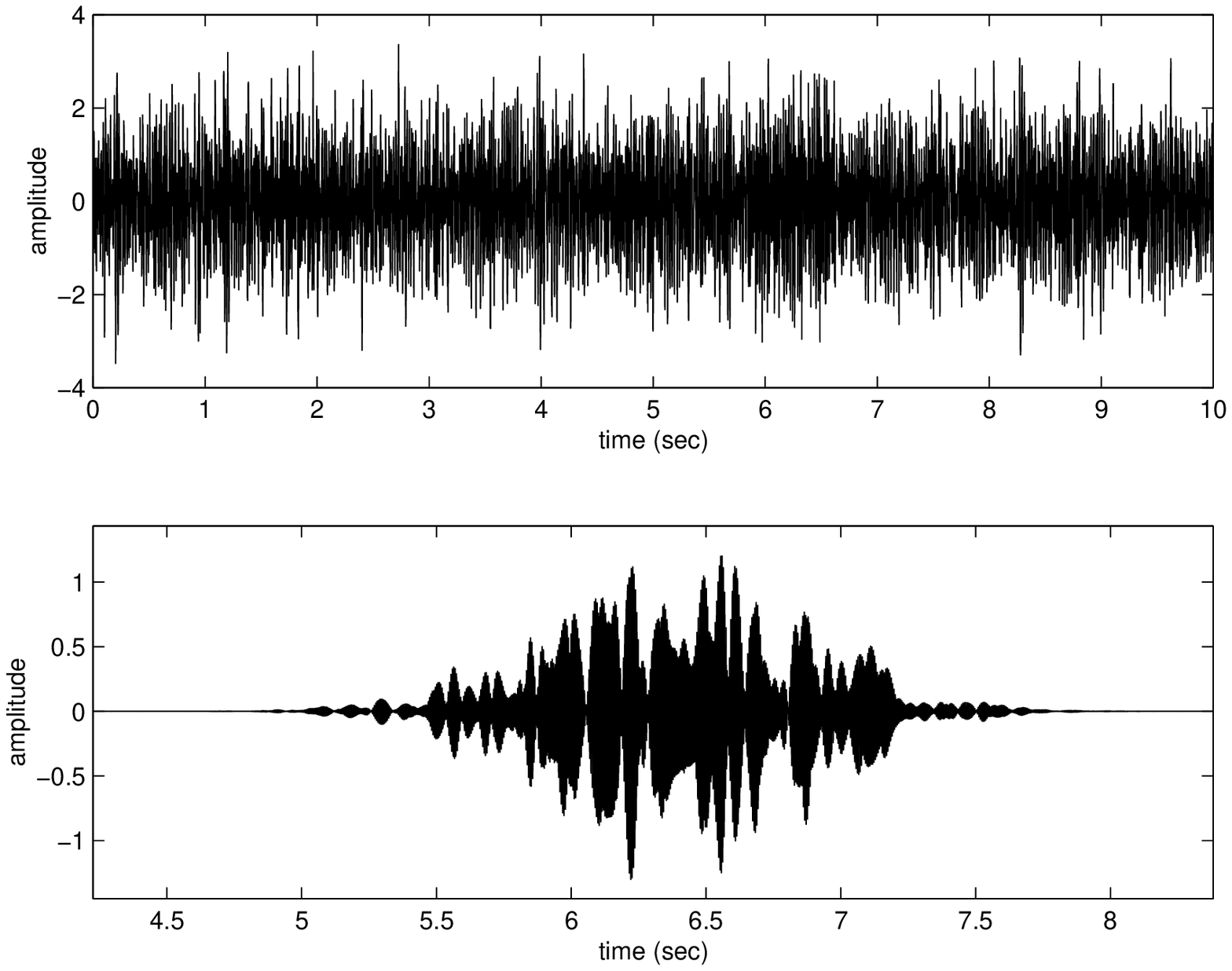,width=3.4in,height=3in}}
\caption{\label{fig6} Sample realizations a narrow band burst 
of type (1) and the corresponding
input data. Top panel: data obtained
 by adding a burst to stationary colored
Gaussian noise.  Bottom: the burst waveform. Here, 
the peak amplitude of the burst is $1.3 \sigma$.}
\end{figure}

\begin{figure}
\centerline{\psfig{file=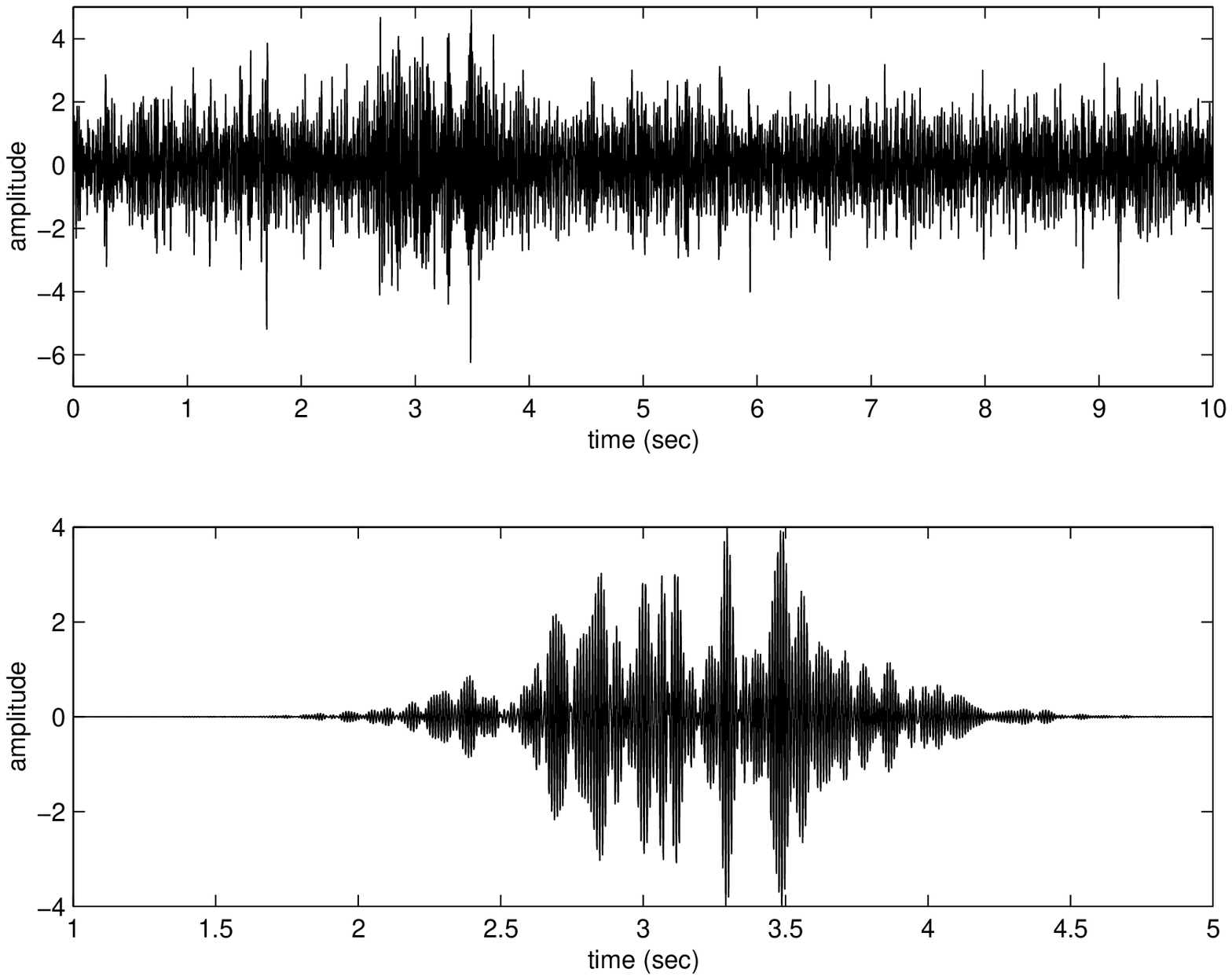,width=3.4in,height=3in}}
\caption{\label{fig7}Sample realizations of a narrow band burst 
of type (2)
and the corresponding
input data. Top panel: data obtained by adding a burst 
to stationary colored
Gaussian noise.  Bottom panel: the burst waveform.
Here, the peak amplitude of the burst is $4.0 \sigma$.}
\end{figure}

In Figs.~\ref{fig6} and \ref{fig7}, we show 
samples of both data and burst (with the peak amplitudes
given in Table~\ref{tableI}) for each of the two
cases described above. Fig.~\ref{fig6} corresponds
to type (1) bursts and  illustrates the fact that
the bursts being detected are not prominent 
enough  to be picked up by ``eye'' .
The burst in Fig.~\ref{fig7}, which is of type (2), is more prominent. 
This is because these bursts lie closer in frequency to the 
``seismic wall''  part of 
the noise curve~(see Fig.~\ref{fig2}) where the variance
of the PSD is higher. 

To better understand the detection efficiency of our test,
it is natural to ask for a comparison with a test that, intuitively,
represents the best we can do. 
Let us suppose that we know {\em a priori} that all
bursts are of type~(2) above and that the
ambient noise is a Gaussian, stationary random process. 
Note that such  
prior information is substantially more than that
used to optimize our test which was a knowledge
of only the burst duration.
 Nonetheless, assuming that such information was
available to us~(and no more), then the following would
be the ideal scheme we should compare our test with.

In the ideal scheme~(similar to~\cite{pallotino+pizzella}),
 we first band pass filter the data ${\bf x}$.
Since we know the bursts are of type~(2),  let the filter
pass band be $W=20$~Hz wide, centered at the frequency $f_c=100$~Hz. 
The output of the filter
is demodulated  and the resulting
quadrature components, say ${\bf X}= \{X_k\}$ and ${\bf Y} =\{Y_k\}$,
$k=1,\,2,\ldots$, are resampled down to a sampling frequency of
$2 W$. The downsampled quadratures are then 
squared and summed to give a time series ${\bf Z}=\{Z_k = X^2_k + Y^2_k\}$.
If any sample of ${\bf Z}$ crosses a threshold $\eta$, we declare that a burst
was present near the location of that sample.

The samples of ${\bf Z}$ should be nearly independent and
 distributed identically. Since the original time series is
a Gaussian random process, this distribution is an exponential.
(Note that the assumption of Gaussianity is essential since 
the central limit theorem does not apply here.) 
The number of samples per hour would be $2 W \times 3600 = 144000$.
For a false alarm rate of one per hour, therefore, the threshold
$\eta$ should be $2.14$. Here, we have used the fact that for
the PSD shown in Fig.~\ref{fig2}, the standard deviation of $Z_k$
turns out to be $0.18$.

Monte Carlo simulations then show that, for obtaining a 
detection probability of 0.8 with the ideal scheme, the
peak amplitude of bursts of type~(2) must be $\simeq 1.5\sigma$,
where $\sigma$ is the standard deviation of the original
time series ${\bf x}$.
From Table~\ref{tableI} we see that, for the same false alarm rate and detection probability,
 our test requires a peak amplitude of 
$4.7\sigma$, a factor of $\sim 3$ higher than that for the ideal test.

\subsection{ The role of the test parameters}
\label{sec3_3}

The test has three adjustable parameters~(c.f.~Sec.~\ref{sec2_2})
 $l_l$,
$\epsilon$ and $l_s$.
The false alarm rate of the test depends 
on the choice of these parameters as does
the power of the test. Here, we empirically explore
 the effect of these parameters on the 
performance of the test.

\subsubsection{Resolution in time and frequency}
The parameter $l_l$, determines the time resolution of
the test. A burst can only be located in time with an accuracy of $\simeq l_l$. 
The duration of a subsegment $l_s$  determines the frequency resolution of the test.
The bin size in frequency domain is simply given by $1/l_s$.

\subsubsection{False alarm rate}
\begin{figure}
\centerline{\psfig{file=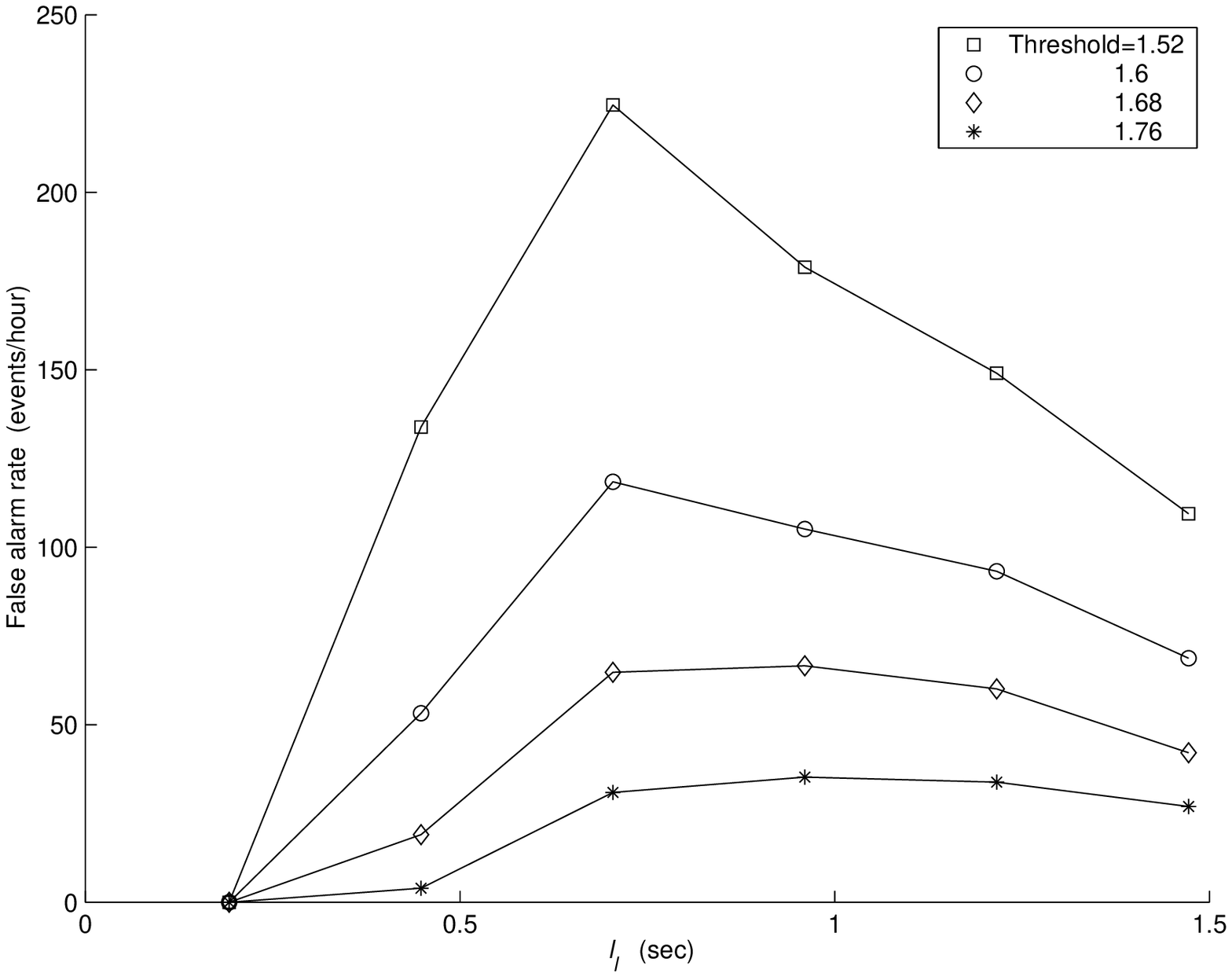,width=3.4in,height=3in}}
\caption{\label{fig8} Effect of $l_l$ ( 
$l_s$ and $\epsilon$ held fixed)
on false alarm rate. The stationary noise used was white Gaussian noise 
($\sigma = 10.0$).}
\end{figure}

 (a) {\em The effect of $l_l$.}
A decrease in $l_l$ reduces the 
 number of samples used in
the {\em t}-test and, hence, should lead to an increase in the 
 false alarm rate. 
Fig.~\ref{fig8} shows the effect of 
$l_l$ on the false alarm rate of the test~($l_s$ and $\epsilon$ held
fixed). It is seen that, for large $l_l$, the trend is indeed as 
expected above but it reverses
below a certain value of $l_l$.
This is probably an effect of the correlation in the 
sequence of {\em t} values~(c.f.~Sec.~\ref{sec2_3}),
though a full understanding requires an analytical
treatment. 
Nonetheless, simulations
establish that this behavior does not significantly affect  the 
robustness of the test. In fact, the parameters
chosen for the simulations in Sec.~\ref{sec3_1}  for
the demonstration of robustness, correspond to
values of $l_l$ on both sides of the change point in Fig.~\ref{fig8}.
Fig.~\ref{fig3} corresponds to a value of $l_l$ that lies
on the left  and Fig.~\ref{fig4} to a value
on the right of the change point and both show that the test
is robust. We have verified this behavior for 
several other cases also.

\begin{figure}
\centerline{\psfig{file=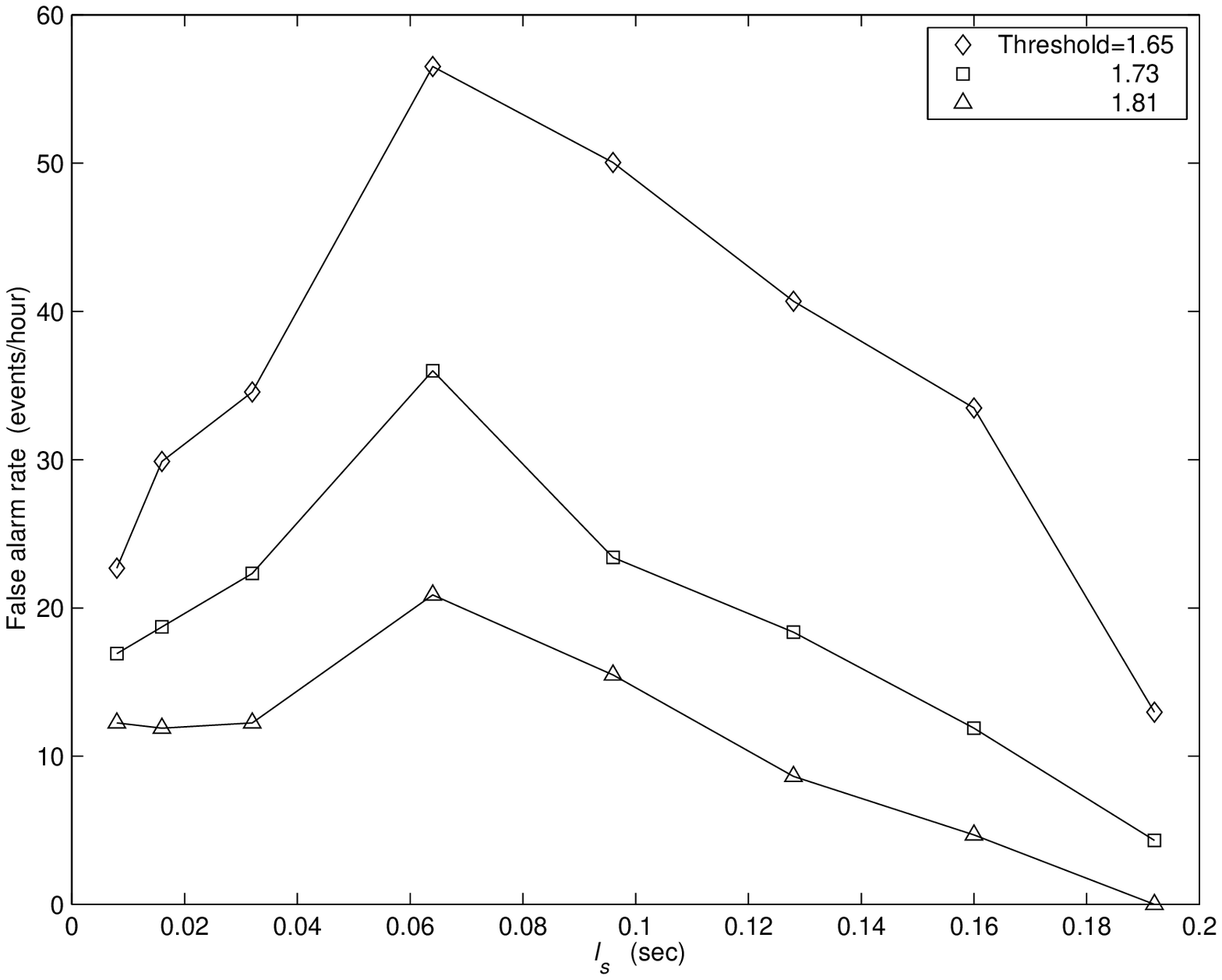,width=3.4in,height=3in}}
\caption{\label{fig9}Effect of  $l_s$ 
($l_l$ and $\epsilon$ held fixed)
on the false alarm rate.
 The stationary noise used was white Gaussian noise ($\sigma = 10.0$).}
\end{figure}

(b) {\em The effect of $l_s$.}
Similarly, the effect of an increase in $l_s$ for a 
fixed $l_l$ is 
expected to increase  the false alarm rate but as in the case
of $l_l$, though this trend is present, it is reversed
above a certain value of $l_s$~(see Fig.~\ref{fig9}). 
  However, simulations verify that
 this does not affect the robustness of the test.

\begin{figure}
\centerline{\psfig{file=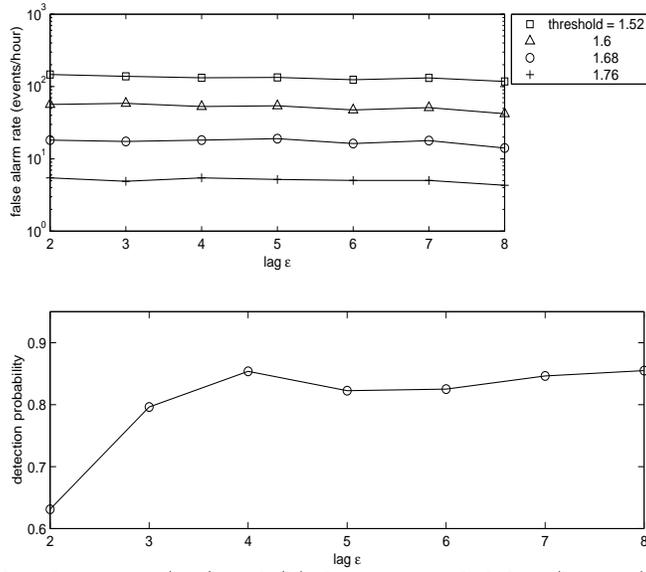,width=3.4in,height=3in}}
\caption{\label{xfig2}
Effect of $\epsilon$ on (a) false alarm 
rate (top) and (b) detection probability (bottom).
 $l_l=0.5$~sec and
$l_s=0.064$~sec for both (a) and (b). The stationary background noise
 used was white Gaussian ($\sigma=10$) for (a) and
colored Gaussian for (b). 
In (b), the sudden drop in detection
probability occurs, as expected, when the burst duration 
(chosen to be $2 l_l$)
becomes comparable to $(\epsilon-1) l_l$ (which is the actual gap). The total
duration of simulated data for (a) was 6.94 hours while the number of 
trials used for (b) was 800.}
\end{figure}

(c) {\em The effect of $\epsilon$.}
The false alarm rate should
 be independent of $\epsilon$ since for {\em stationary} noise
 it does not matter which two segments  are compared in the {\em t}-test.
This agrees with actual simulation results as shown in Fig.~\ref{xfig2}(a).

\subsubsection{Detection probability}
\begin{figure}
\centerline{\psfig{file=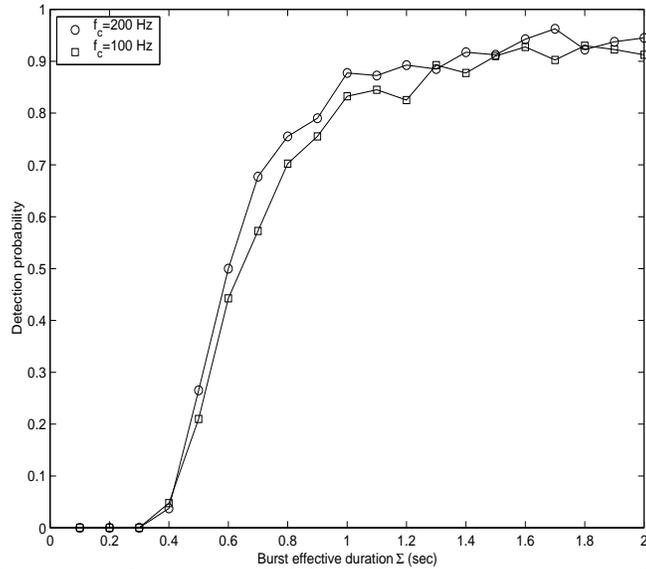,width=3.4in,height=3in}}
\caption{\label{fig10} The effect of burst 
duration on detection probability. The test 
parameters were fixed at $l_l=0.5$~sec, $l_s=0.064$~sec and $\epsilon =5$. 
The burst peak amplitude for the type~(1) bursts ($f_c=200$~Hz)
was $1.6\sigma$ while it was $4.7\sigma$
for the type~(2) bursts ($f_c=100$~Hz).
 The false alarm rate was fixed at 1 false event/hour.}
\end{figure}

(a) {\em The effect of $l_l$.}
 When $l_l$
is significantly larger than the burst duration, only
a few of the subsegments in  the segment containing 
the burst will have a  
distribution which is different from the stationary case.
The periodograms of such subsegments will appear as outliers 
in an otherwise normal sample and the {\em t}-test, which is  
unsuitable for such cases, will not be able to detect them. 
Therefore, as the burst duration falls below
$l_l$, the detection probability should decrease.
The effect of $l_l$ on detection probability 
should be independent of the frequency band 
in which the burst is localized since $l_l$
only governs the number of subsegments  
over which the burst is spread. Both of
the above effects are observed, as shown in Fig.~\ref{fig10}.
Thus, to optimize the test, {\em the only prior knowledge
required} is the duration of the bursts which are to
be detected.

(b) {\em The effect of $l_s$.}
Decreasing $l_s$ will increase $N$, the number of
samples used in the {\em t}-test, which 
would increase the detection probability.
However, $l_s$ should not be reduced
indiscriminately (see Sec.~\ref{subsub_misc}).

(c) {\em The lag $\epsilon$.}
 As long
as the burst durations are smaller than $\epsilon$,
a change in $\epsilon$ should not affect the 
power of the test. 
This is indeed observed in our
simulations, an example of which is shown
  in Fig.~\ref{xfig2}(b). In Fig.~\ref{fig10}, we
kept $\epsilon$ large enough that the effect of
$\epsilon$ on burst  detectability did not 
get entangled with that of $l_l$.

\subsubsection{Miscellaneous}
\label{subsub_misc}
Reducing $l_s$ to the point that each
subsegment has only one sample is simply
equivalent to monitoring changes in the variance
of the input time series. This is because a one
point DFT is simply the sample itself and the
periodogram is, therefore, just the square of the 
sample. {\em Thus, a test for change in variance
is a special case of the present test.}

However, under some circumstances, an
indiscriminate reduction in $l_s$ can have
adverse effects. For instance, suppose the
ambient noise PSD is such that the power in some
frequency region is much greater than the power 
elsewhere~(see Fig.~\ref{fig2} for an example) and all the bursts  occur in
the low power region.  Since reduction in $l_s$
decreases frequency resolution, the low power 
region will be completely masked by the high power
one for sufficiently small $l_s$. This will then
make the detection of the bursts more difficult.
A related issue  is that of narrowband noise
contamination which is discussed in more detail
in Sec.~\ref{sec5}.

A very long lag would allow the 
detection of long time scale non-stationarity
such as an abrupt change in the variance 
from one fixed value to another. However, for such abrupt
long lasting changes, there  exist
better methods of detection~\cite{cusum}.

\subsection{Computational Cost}
\label{sec3_4}

In estimating the computational cost of this test,
it is helpful to divide the total number of floating point 
operations~(additions, subtractions, multiplications) required into
two parts : (a) Deterministic and (b) Stochastic.

(a) {\em Deterministic}. This is the part involving
the generation of the raw image (c.f.~Sec.~\ref{sec2_2}).
The number of operations required is completely determined
by the parameters $l_l$, $l_s$ and the sampling frequency of 
the data $f_s$. A breakup of the steps involved in this 
part and the respective number of
operations involved is as follows.

For each column of the image : (1) Two sets of Fast Fourier Transforms (FFTs)
have to be computed, each set having $N=l_l/l_s$ FFTs with each FFT 
involving $n=l_s f_s$ time samples. Therefore, the number of operations
involved is $2\times N\times 3 n \log_2 n$. (2) The modulus squared
of only the positive frequency FFT amplitudes are computed for each 
subsegment leading to $2 N\times 3\times (n/2)$ operations (the 
factor 3 comes from squaring and adding the real and imaginary parts).
(3) For each frequency,
the sample mean ($2N+1$ operations)
and variances ($3N+1$ operations)  are computed 
followed by 4 operations to construct the {\em t}-statistic.
Thus, total number of operations involved is $(5N+8)n/2 $.
(4) Finally, for each frequency,
the {\em t}-statistic is compared to a threshold,
involving $n/2$ operations in all. Adding up all the steps and
dividing the total number of operations by $l_l$ gives the 
computing speed  required in order to generate the image 
online : $\left(6 \log_2 n + 9/2 N+ 11/2 \right)f_s $.
As an example, for $f_s=5000$~Hz, $l_l=0.5$~sec and
$l_s = 0.064$~sec, the required computing speed is
$0.28$~MFlops. Thus, generating the raw image is computationally trivial
by the standards of modern day computing.

(b) {\em Stochastic}. This is the part involving the
application of the veto to the raw image (c.f.~Sec.~\ref{sec2_2}).
Since the number of black pixels in the image after 
thresholding as well as the size of the black-pixel patches
are random variables, the number of operations involved 
in this part is also a random variable. One expects, however,
that for low false alarm rates, the computational cost of
this part will be much smaller than that of the deterministic
part since clusters would only occur sparsely in the image.

The simplest way to estimate the computational load
because of the stochastic part
is via Monte Carlo simulations in which the number of
operations involved in the stochastic part are explicitly
counted within the code itself. In Table~\ref{tableII},
we present the number of floating point operations 
incurred in the stochastic part, as a fraction of the 
total number of operations incurred in the deterministic 
part, over a wide range of false alarm rates for stationary input noise.
(To generate Table~\ref{tableII},
the test parameters used were $l_l=0.5$~sec, 
$l_s=0.064$~sec and $\epsilon=3$. The sampling frequency of the 
input data was 1000~Hz, each realization being 20~sec long.
The operations were counted over 200 trials.)

From Table~\ref{tableII}, we see that even when the
false alarm rate is as high as 50 events/hour, the
time spent in the stochastic part is negligible 
compared to that involved in generating the raw
image itself. The computational cost of generating
the image itself~(the deterministic part) is 
quite low as shown above. Hence, overall, the test
can be implemented without significant computational
costs.

\section{Discussion}
\label{sec5}

A test for the detection of non-stationarity is presented
which has the important property of being robust.
This allows the test to be used on data without the need to
first characterize
the data statistically. 

The main results of this work are (i) the demonstration, using 
Monte Carlo simulations, of the insensitivity of the false
alarm rate at a given threshold to the statistical nature of 
the data being analyzed, and (ii) application of the test to
the detection of different types of
bursts which showed that the test can detect fairly weak
 bursts. For instance, as shown in Table~\ref{tableI}, the test could
detect 80\% of narrowband bursts, each 
located  within a band of 20~Hz centered at 200~Hz,
that were added to Gaussian noise with a PSD such as that of LIGO-I 
when the peak amplitude of the bursts was only
$1.6\times \mbox{r.m.s.~of background noise}$ and the false alarm rate 
for the test was 1~event/hour. 

We did not catalog the
false alarm rate or detection probability
 for a large number of
cases since real applications will almost always fall outside any
such catalog. Instead, for false alarm rate, we chose a rather extreme
range for the types of stationary noise so that a bound on the robustness
could be obtained. While, for detection probability, our main aim was to
demonstrate that, given its robustness, the test performs quite well in
realistic situations. When applying the test to a particular data set,
the appropriate false alarm versus threshold curve can be obtained easily
using a single Monte Carlo simulation. Almost always, the experimenter
has some prior idea of the {\em range} of burst durations he/she is 
interested in and therefore can choose the set of test parameters appropriately.
This would be necessary for any test of non-stationarity, and not 
particularly the present one, since non-stationarity can take many forms.
A more general approach would require understanding the test
analytically. This work is in progress.
 
Though we mentioned the problem of narrow band noise 
(c.f.~Sec.~\ref{sec3_3})  
it was not addressed in detail.
This is because this is an issue that is fundamental to
all tests for transient non-stationarity and not specific to
the present test alone. Narrow band noise, such as power
supply interference at 60~Hz and its harmonics or the thermal
noise associated with the violin modes of suspension wires,
appear non-stationary on timescales much shorter than their correlation length.
Thus, if a narrow band noise component has
 significant power, the frequency band (max[frequency 
resolution, line width])
containing it
will appear non-stationary to any test that searches for 
short duration transients.
On the other hand,  {\em steady} narrowband signals
in the data can suppress the detection of non-stationarity that happens
to lie close to them in frequency. This is because detection
of short bursts implies an increase in time resolution and, correspondingly,
a decrease in frequency resolution. Thus if the narrowband signals are strong,
they can make the frequency bins containing them appear stationary. 

This problem can be addressed in several ways. A preliminary look at the
PSD can tell us about the frequency bands where narrowband interference
is severe and the output of the test in those bands can be discarded
from further analysis. Another way could be to decrease the time resolution
sufficiently though at the cost of losing short bursts. A more direct and 
effective approach would be to pass the data through time domain filters
that notch the offending frequencies. Such filters could also be  made
adaptive so that the frequencies can be tracked in time~\cite{tone_removal}.
Further work is in progress on this issue.
\section*{Acknowledgement}
I am very grateful to Albert~Lazzarini for extremely useful
comments, criticisms and suggestions which led to a significant
improvement in the work. 
 I thank Albert~Lazzarini and Daniel~Sigg for help
in obtaining information about auxiliary channels in LIGO.
I thank  Eric-Chassande Motin for discussions that led to some interesting
ideas for the future.
 Discussions with Gabriela~Gonzalez
 and L.~S.~Finn were helpful. I thank L.~S.~Finn for suggestions
regarding the text. 
This work was supported by National Science Foundation awards PHY
98-00111 and PHY 99-96213 to The Pennsylvania State University.

\begin{appendix}
\section*{Algorithm of the test}
\label{app1}
\subsection{Notation}

We present, first, some of the notation that will be used
in the following. 
The time series to be analyzed will be denoted by $ {\bf x}$,
where $ {\bf x}$ is a sequence of
real numbers.
We will need to
divide $ {\bf x}$  into disjoint segments, without
gaps, with all segments having the same duration $l_l$. A segment of length $l_l$ will 
be denoted by $ {\bf y}^{(j)}$, where $j$ stands for segment
number $j$.  

Each segment $ {\bf y}^{(j)}$
will need to be further subdivided into disjoint segments, again without 
gaps, with all subsegments having the same duration $l_s$. The $k^{\rm th}$
such sub-segment in the segment $ {\bf y}^{(j)}$ will be denoted 
by $ {\bf z}^{(j,k)}$.

The  {\em periodogram} of a time series is defined to be the squared
modulus of its DFT.
That is, if $\widehat{\bf u}$ is the DFT of some time series
$ {\bf u}$~(consisting of $m$ samples), then the $q^{\rm th}$
frequency component  $\widehat{u}_q$ of $\widehat{\bf u}$ is given by,
\begin{equation}
\widehat{u}_q := 
\sum_{p=1}^{m} u_p \exp\left(
{2 \pi i (q-1) (p-1)/m}\right)\;.
\end{equation}
where $q=\{1,\ldots,m\}$. The periodogram $\{S_q\}$~($q=\{1,\ldots,m\}$)
is defined by
\begin{equation}
S_q := |\widehat{u}_q|^2\;. \label{pdg}
\end{equation}
To reduce the aliasing of high frequency power on to lower frequencies, 
it is common to compute the periodogram after modifying ${\bf u}$
by multiplying it with a {\em window}
function $ {\bf w}$ : $u_p$ $\rightarrow$ 
$ u_p w_p$. 
The definition of the periodogram is modified in this case to,
\begin{equation}
S_q := \frac{1}{\parallel {\bf w} \parallel} |\widetilde{u}_q|^2
\label{mod_psd}
\end{equation}
where $\parallel {\bf w} \parallel$ stands for the Euclidean norm of the
window function and $\widetilde{u}_q$ is the $q^{\rm th}$ frequency 
component of
the DFT of the windowed sequence. Before windowing, we also subtract the sample mean
of the sequence from each sample in the sequence.
In the following, all periodograms are obtained as defined in 
Eq.~(\ref{mod_psd}) after subtraction of the sample mean followed by windowing.
the window function is chosen to be the
symmetric Hanning window of the same length as the input time series ${\bf u}$.
We denote the periodogram of $ {\bf z}^{(j,k)}$ by ${\bf S}^{(j,k)}$,
with its $q^{\rm th}$ frequency component denoted by $S^{(j,k)}_q$.

\subsection{Algorithm : The first stage}

We will now state the algorithm of the test. 
First, the values of the
free parameters of the test $l_l$, $l_s$ and $\epsilon$
 are set. 
Then the following loop
is executed.
\begin{enumerate}
\item Starting with $j=1$, take segments ${\bf y}^{(j)}$ and
${\bf y}^{(j+\epsilon)}$ from the detector output ${\bf x}$.
The loop index is $j$.
\item Subdivide each of the above segments into equal number of 
subsegments ${\bf z}^{(j,k)}$ and ${\bf z}^{(j+\epsilon,k)}$, $k=1,\ldots, 
{\rm floor}(l_l/l_s)$, where the floor function returns the integer part
of its argument.  Let $N = {\rm floor}(l_l/l_s)$.
\item Compute the sets $\{ {\bf S}^{(j,k)}\}$ and $\{{\bf S}^{(j+\epsilon,k)}
\}$  with $k$ as the running index.
Thus, each of the two sets contains  $N$ periodograms. 
\item For each frequency component, compute the {\em sample} means 
and variances of the two sets. Let the sample means at the $q^{\rm th}$ 
frequency 
component  be denoted by
$\mu_q^{(j)}$ and $\mu_q^{(j+\epsilon)}$ for $\{ {\bf S}^{(j,k)}\}$
and $\{{\bf S}^{(j+\epsilon,k)}
\}$ respectively. Then,
\begin{eqnarray}
\mu_q^{(j)} & = & N^{-1}\sum_{k=1}^{N} S^{(j,k)}_q \nonumber \;,\\
\mu_q^{(j+\epsilon)} & = & N^{-1}\sum_{k=1}^{N} S^{(j+\epsilon,k)}_q \;.
\nonumber
\end{eqnarray}
Similarly, let the standard deviations be denoted by $\sigma^{(j)}_q$
and $\sigma^{(j+\epsilon)}_q$,
\begin{eqnarray}
\left(\sigma_q^{(j)}\right)^2 & = & 
(N-1)^{-1}\sum_{k=1}^{N} \left(S^{(j,k)}_q -\mu_q^{(j)}\right)^2
\nonumber \;,\\
\left(\sigma_q^{(j+\epsilon)}\right)^2& = & 
(N-1)^{-1}\sum_{k=1}^{N} \left(S^{(j+\epsilon,k)}_q -
\mu_q^{(j+\epsilon)}\right)^2\;,
\nonumber
\end{eqnarray}
where we have used the unbiased estimator of variance (the biased
estimator has $N$ in the denominator instead of $N-1$).
\item Compute $t_q^{(j)}$, the value of the {\em t}-statistic for the 
$q^{\rm th}$ frequency component,
\begin{equation}
t_q^{(j)} := \sqrt{N}\, \frac{\mu_q^{(j+\epsilon)}-\mu_q^{(j)}}{
\left[ \left(\sigma_q^{(j)}\right)^2 +
\left(\sigma_q^{(j+\epsilon)}\right)^2\right]^{1/2}}\;.
\label{tqj}
\end{equation}
\end{enumerate}
Let ${\bf T}$ be a matrix with  ${\bf T}_{qj}= |t_q^{(j)}|$, 
$q$ and $j$ being the row and column indices respectively.
For every pass through the loop described above, 
a column of ${\bf T}$ is produced. 

Let the threshold for the {\em t}-test be 
 $\eta$. Set all elements of ${\bf T}$ that are below
$\eta$ to zero and set all elements above $\eta$ to a fixed value $t_0$. 
We denote the resulting matrix by the same symbol ${\bf T}$. This should not
cause any confusion since we will mostly require the thresholded form of
${\bf T}$ in the following.

The matrix ${\bf T}$ can also be 
visualized~(see Fig.~\ref{fig1}) as a two dimensional
 image composed of a rectangular array of {\em pixels}~(picture elements)
with the same dimension as ${\bf T}$. 
We can imagine that the
pixels for which the corresponding
matrix elements crossed $\eta$ are colored black and those that did not cross
are colored white.
 We call the black pixels  {\em b}-pixels and the white ones 
{\em w}-pixels.

\subsection{Algorithm : The second stage}

\begin{figure}
\centerline{\psfig{file=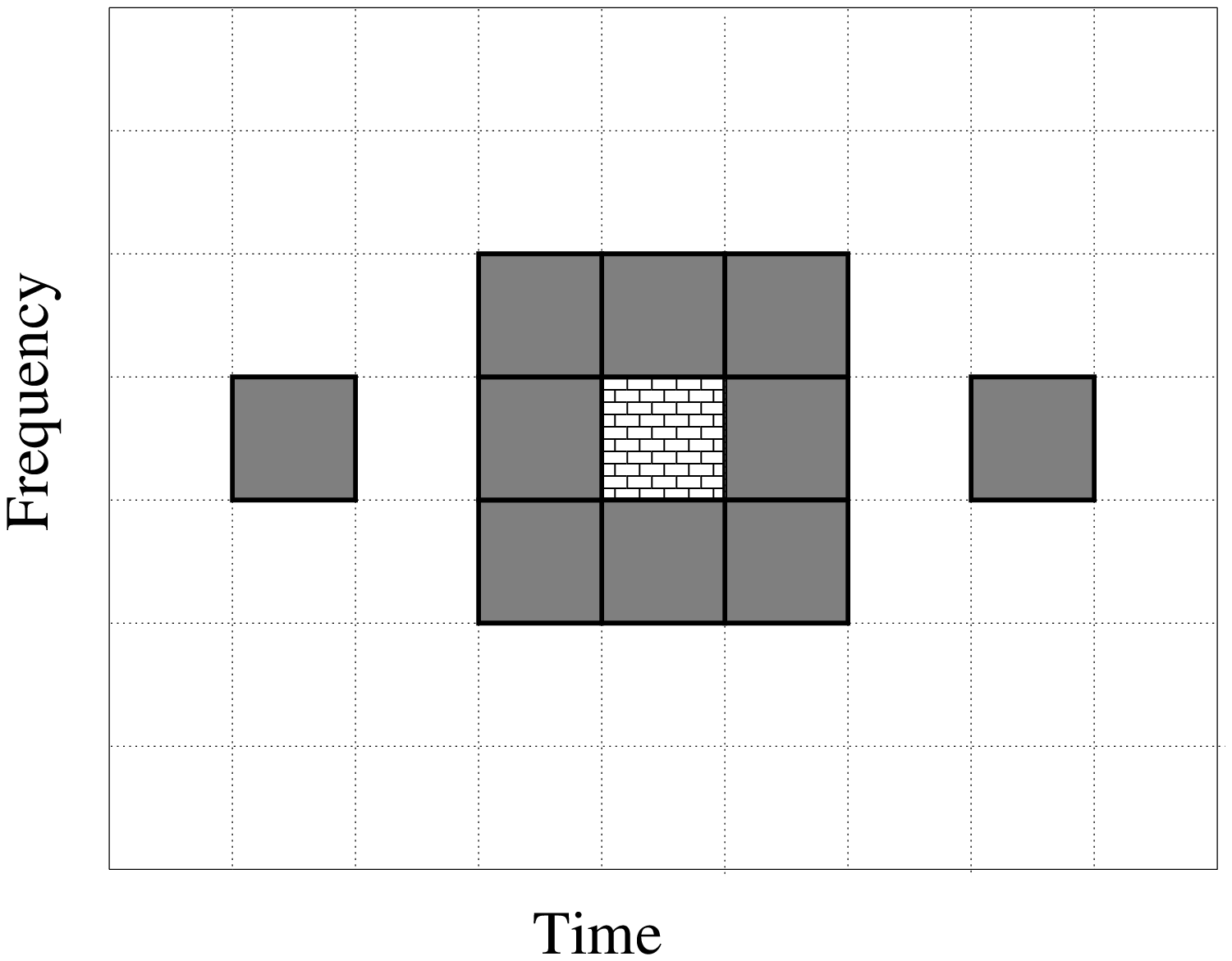,width=3.4in,height=3in}}
\caption{\label{fig11}  Nearest neighbors (gray) of a
pixel (brick-wall pattern).
The gray pixels which touch the central pixel 
are its {\em contacting} nearest 
neighbors while the two that do not are its {\em non-contacting } nearest
neighbors. ($\epsilon = 3$ here.)}
\end{figure}

A burst will appear in the image ${\bf T}$
as a {\em cluster} of {\em b}-pixels.
 In order to define a cluster we first delineate 
the set of pixels which form the {\em nearest
neighbors} to any given pixel. The nearest neighbor
of a pixel with $q$ as the row and
$j$ as the column index is a pixel with
row index $q^\prime$ and column index $j^\prime$ such that
(i) $q^\prime \in \{ q,\,q\pm 1\}$ and
 $j^\prime \in \{ j,\,j\pm 1\}$ or (ii) $q^\prime = q$ and 
$j^\prime \in \{j+\epsilon, j-\epsilon\}$. 
We call the set of nearest neighbors of type (i) as 
{\em contacting} and those of type (ii) as {\em non-contacting}.
Fig.~\ref{fig11}
shows the set of nearest neighbors of a pixel.
We can now define a cluster of {\em b}-pixels as 
a set of {\em b}-pixels such that (i) each member of this
set has at least one other member as its nearest neighbor,
and (ii) at least one member of the cluster has another member 
 as a non-contacting nearest neighbor.

The next step in the 
algorithm is the identification of a cluster of {\em b}-pixels
in the image ${\bf T}$. In our code, we proceed as follows.
Make a list of 
all {\em b}-pixels in the image ${\bf T}$~(the ordering of the list is immaterial). 
Let this list be called ${\bf L}$. We define two more symbols :
\begin{description}
\item{(i)} ${\bf L}_{\rm sub}$ is a proper subset of ${\bf L}$ such that 
any two elements $a\in {\bf L}_{\rm sub}$ and $b\in {\bf L}_{\rm sub}$, 
there exist elements $\{c,\, d,\,\ldots,\,h\}\in {\bf L}_{\rm sub}$ such
that $c$ is a contacting nearest neighbor of $a$, $d$ is a
contacting nearest neighbor of $c$ and so on till $h$ which is also a 
contacting nearest neighbor of $b$. That is, starting from any one element
we can reach any other by ``stepping'' through a chain
of members. Essentially, ${\bf L}_{\rm sub}$
is, roughly speaking, an unbroken patch of {\em b}-pixels.
\item{(ii)} ${\bf L}_{\rm sub}^\prime$ is the complement of ${\bf L}_{\rm sub}$
in ${\bf L}$.
\end{description}
In the algorithm below, it is understood that
when an element is added or removed from ${\bf L}_{\rm sub}$,
the new set is always renamed as ${\bf L}_{\rm sub}$.
Similarly, the complement of the new ${\bf L}_{\rm sub}$
is always denoted by ${\bf L}_{\rm sub}^\prime$.

The steps in the algorithm are as follows. (Parenthesized statements
are comments.)
\begin{enumerate}
\item \label{step1}   For
each member of ${\bf L}_{\rm sub}$, 
search for {\em contacting} nearest neighbors in ${\bf L}_{\rm sub}^\prime$. 

\item \label{step2} if found  add them to ${\bf L}_{\rm sub}$.
 If not, go to step~\ref{step4}.

(To obtain 
${\bf L}_{\rm sub}$ starting from the null set: 
take the first element, which we call the
{\em seed} element, of ${\bf L}$ as ${\bf L}_{\rm sub}$
 and go to step~\ref{step1}.) 
\item \label{step3}
 Update
${\bf L}_{\rm sub}^\prime$. Go to step~\ref{step1}.

\item \label{step4}
Check if any 
element of ${\bf L}_{\rm sub}$ has a {\em non-contacting} nearest
neighbor in ${\bf L}_{\rm sub}^\prime$. 

(This and the following steps check  whether ${\bf L}_{\rm sub}$ qualifies as a 
cluster according to our definition.) 

\item \label{step5}If none are found, go to step~\ref{step7}. Otherwise,
 take the first non-contacting nearest neighbor as a new seed element
and construct a subset 
$\widetilde{\bf L}_{\rm sub}$ following 
step~\ref{step1} to step~\ref{step3} 
(temporarily rename ${\bf L}$ by ${\bf L}_{\rm sub}^\prime$,
${\bf L}_{\rm sub}$ by $\widetilde{\bf L}_{\rm sub}$
and ${\bf L}_{\rm sub}^\prime$ by $\widetilde{\bf L}_{\rm sub}^\prime$ 
in those
steps). Add $\widetilde{\bf L}_{\rm sub}^\prime$ to ${\bf L}_{\rm sub}$
and set a flag ${\cal B}$ that ${\bf L}_{\rm sub}$ is a cluster.
 
\item \label{step6}
Repeat step~\ref{step4}. 

\item  \label{step7} Rename ${\bf L}^\prime_{\rm sub}$  
as ${\bf L}$. If flag ${\cal B}$ was set, save ${\bf L}_{\rm sub}$.
 Go to step~\ref{step1} again (until 
not more than one {\em b}-pixel is left in ${\bf L}$).
\end{enumerate}
The  above algorithm is easy to implement in 
softwares such as {\sc MATLAB}~\cite{mathworks} or 
{\sc MATHEMATICA}~\cite{wolfram}
which have inbuilt routines for set operations. 
We use {\sc MATLAB} for our implementation. The 
actual code can, of course, be optimized significantly.
For instance, in step~\ref{step1} the search can be
confined to only the most recent set of elements added to ${\bf L}_{\rm sub}$.
\end{appendix}
\newpage
\begin{table}
\caption{\label{tableI}
Burst peak amplitude, in multiples of the background noise r.m.s.,
required for a detection probability of 0.8. 
The threshold $\eta$ corresponding
to a particular false alarm rate is given in parantheses below it.}
\begin{tabular}{|lllll|}
Burst  & \multicolumn{4}{c}{False alarm rate}  \\
type& \multicolumn{4}{c}{(number of events/$x$ hours)}\\
& 2 /hr  & 1 /hr  & 1 /2hrs
 & 1 /3hrs \\
 &($\eta =1.8$)&(1.84)& (1.875) &(1.9)\\
\hline
(1) & 1.3& 1.6& 1.8 & 2.3\\
(2) & 4.0& 4.7&5.8&6.4 \\ 
\end{tabular}
\end{table}

\begin{table}
\caption{\label{tableII} Number of operations involved in the
stochastic part as a fraction of the number of operations required
in the deterministic part. 
Numbers in parentheses are the false alarm rates corresponding
to the respective thresholds $\eta$.}
\begin{tabular}{|cccc|}
$\eta=1.57$ & 1.65 & 1.7 & 1.8\\
(50 events/hour) & (20/hr) & (10/hr) & (2/hr)\\
\hline
$0.46\times 10^{-3}$ & $0.23\times 10^{-3}$ & $0.14\times 10^{-3}$ & 
$0.04\times 10^{-3}$
\end{tabular}
\end{table}
\newpage

\begin{references}
\bibitem{exponential} An exponential probability density function is given by 
$p(x) = \exp(-x/\sigma)/\sigma$, where $\sigma$ is both the mean and 
standard deviation of $p(x)$. For instance, the sum of the squares of 
two independent
and identically distributed Gaussian random variables is distributed 
as an exponential.
\end{references}
\end{document}